\documentclass[prl,twocolumn,superscriptaddress,showpacs]{revtex4}

\usepackage{graphicx}
\usepackage{amssymb,amsmath,color}

\newcommand{\ket}[1]{|#1\rangle}

\begin{document}

\title{14-qubit entanglement: creation and coherence}

\date{\today}

\author{Thomas~Monz}
\affiliation{Institut f\"ur Experimentalphysik, Universit\"at
Innsbruck, Technikerstr. 25, A-6020 Innsbruck, Austria}
\email{Thomas.Monz@uibk.ac.at}

\author{Philipp~Schindler}
\affiliation{Institut f\"ur Experimentalphysik, Universit\"at
Innsbruck, Technikerstr. 25, A-6020 Innsbruck, Austria}

\author{Julio~T.~Barreiro}
\affiliation{Institut f\"ur Experimentalphysik, Universit\"at
Innsbruck, Technikerstr. 25, A-6020 Innsbruck, Austria}

\author{Michael~Chwalla}
\affiliation{Institut f\"ur Experimentalphysik, Universit\"at
Innsbruck, Technikerstr. 25, A-6020 Innsbruck, Austria}

\author{Daniel~Nigg}
\affiliation{Institut f\"ur Experimentalphysik, Universit\"at
Innsbruck, Technikerstr. 25, A-6020 Innsbruck, Austria}

\author{William~A.~Coish}
\affiliation{Institute for Quantum Computing and Department
of Physics and Astronomy, University of Waterloo, Waterloo,
ON, N2L 3G1, Canada}
\affiliation{Department of Physics, McGill University, Montreal,
Quebec, Canada H3A 2T8}

\author{Maximilian~Harlander}
\affiliation{Institut f\"ur Experimentalphysik, Universit\"at
Innsbruck, Technikerstr. 25, A-6020 Innsbruck, Austria}

\author{Wolfgang~H\"ansel}
\affiliation{Institut f\"ur Quantenoptik und Quanteninformation,
\"Osterreichische Akademie der Wissenschaften, Otto-Hittmair-Platz
1, A-6020 Innsbruck, Austria}

\author{Markus~Hennrich}
\affiliation{Institut f\"ur Experimentalphysik, Universit\"at
Innsbruck, Technikerstr. 25, A-6020 Innsbruck, Austria}
\email{Markus.Hennrich@uibk.ac.at}

\author{Rainer~Blatt}
\affiliation{Institut f\"ur Experimentalphysik, Universit\"at
Innsbruck, Technikerstr. 25, A-6020 Innsbruck, Austria}
\affiliation{Institut f\"ur Quantenoptik und Quanteninformation,
\"Osterreichische Akademie der Wissenschaften, Otto-Hittmair-Platz
1, A-6020 Innsbruck, Austria}


\begin{abstract}
  We report the creation of Greenberger-Horne-Zeilinger states with up
  to 14 qubits. By investigating the coherence of up to 8 ions over
  time, we observe a decay proportional to the square of the number of
  qubits. The observed decay agrees with a theoretical model which
  assumes a system affected by correlated, Gaussian phase noise. This
  model holds for the majority of current experimental systems
  developed towards quantum computation and quantum metrology.
\end{abstract}

\maketitle

Quantum states can show non-classical properties, for example, their
superposition allows for (classically) counter-intuitive situations
such as a particle being in two places at the same time. Entanglement
can extend this paradox even further, e.g., the state of one subsystem
can be affected by a measurement on another subsystem without any
apparent interaction~\cite{epr_paper}. These concepts, although
experimentally frequently verified, contrast with our classical
perception and lead to several questions. Is there a transition from a
quantum to a classical regime? Under which conditions does that
transition take place? And why? The creation of large-scale
multi-particle entangled quantum states and the investigation of their
decay towards classicality may provide a better understanding of this
transition~\cite{macro_realism,lintrapBE, haroche_decay,
  wineland_decay}.

Usually, decoherence mechanisms are used to describe the evolution of
a quantum system into the classical regime. One prominent example is
the spontaneous decay of the excited state of an atom. In a collection
of atoms, the decay of each would be expected to be independent of the
others. Therefore, the number of decay processes in a fixed time
window would intuitively be proportional to the number of excited
atoms. This assumption, however, can be inaccurate. Decoherence
effects can act collectively and produce ``superradiance'', a regime
in which the rate of spontaneous decay is proportional to the square
of the number of excited atoms~\cite{dicke_radiation}. Such collective
decoherence can also occur in multi-qubit registers, an effect known
as ``superdecoherence''~\cite{superdecoherence}. This particularly
applies to most currently used qubits which are encoded in
energetically non-degenerate states. In these systems, a phase
reference (PR) is required to perform coherent operations on a quantum
register. Noise in this PR thus collectively affects the quantum
register.

In the following we introduce a model describing a quantum register in
the presence of correlated phase noise. More specifically, we
investigate $N$-qubit Greenberger-Horne-Zeilinger (GHZ) states of the
form $\ket{\psi(0)} = \frac{1}{\sqrt{2}}(\ket{0 \ldots 0}+\ket{1
  \ldots 1})$. These states are the archetype of multi-particle
entanglement and play an important role in the field of quantum
metrology~\cite{quantum_metrology} for quantum-mechanically enhanced
sensors. This special quantum state, however, has only been generated
with up to 6 particles so far~\cite{ghz_6_didi, hyper_ent_pan}.
Employing up to 8 genuinely multi-particle-entangled ion-qubits in a
GHZ state, we predict and verify the presence of superdecoherence
which scales quadratically with the number of qubits $N$.
In general, any system experiencing correlated phase noise is affected by this
accelerated GHZ-state decoherence.


\begin{table*}
  \caption{Populations, coherence, and fidelity with a $N$-qubit GHZ-state of
    experimentally prepared states. Entanglement criteria supported by
    $\sigma$ standard deviations. All errors in parenthesis, one standard
    deviation.}
\begin{ruledtabular}
\begin{tabular}{lccccccccc}
  Number of ions & 2 & 3 & 4 & 5 & 6 & 8 & 10 & 12 & 14\\\hline
  Populations, \% & 99.50(7) & 97.6(2) & 97.5(2) & 96.0(4) & 91.6(4) & 84.7(4) & 67.0(8) & 53.3(9) & 56.2(11)  \\
  Coherence, \% & 97.8(3) & 96.5(6) & 93.9(5) & 92.9(8) & 86.8(8) & 78.7(7) & 58.2(9) & 41.6(10) & 45.4(13)  \\
  Fidelity, \% & 98.6(2) & 97.0(3) & 95.7(3) & 94.4(5) & 89.2(4) & 81.7(4) & 62.6(6) & 47.4(7) & 50.8(9)  \\
  Distillability criterion~\cite{ghz_threshold_wolfgang}, $\sigma$ & 283 & 151 & 181 & 100 & 95 & 96 & 40 & 18 & 17 \\
  Entanglement criterion~\cite{ghz_crit_otfried}, $\sigma$ & 265 & 143 & 167 & 101 & 96 & 92 & 25 & -6 & 0.7
\end{tabular}
\end{ruledtabular}
\end{table*}

We model collective phase fluctuations acting on the quantum register
with a Hamiltonian of the form
$H_{\mathrm{noise}} = \frac{\Delta E (t)}{2} \sum_{k=1}^{N} \sigma_z^{(k)}$
where $\Delta E (t)$ denotes the strength of the fluctuations, and
$\sigma_z^{(k)}$ a phase flip on the k-th ion. Under this
Hamiltonian, the initial state of the system $|\psi(0) \rangle$
evolves into $|\psi(t)\rangle = \exp(-\frac{i}{\hbar} \int_0^t
\mathrm{d}\tau~H_\mathrm{noise}(\tau)) |\psi(0)\rangle$. As a measure
of state preservation, we use the fidelity
$F(t)=\overline{|\langle\psi(0)|\psi(t)\rangle|^2}$, where the bar
refers to an average over all realizations of random phase
fluctuations. The decay of this fidelity can be conveniently described
by
\begin{equation}
F(t) = \frac{1}{2} (1+\exp(-2 \epsilon (N,t))),\nonumber
\end{equation}
where the effective error probability for a stationary Gaussian random
process is derived to be
\begin{equation}
\epsilon (N,t) = N^2 \frac{1}{2
  \hbar^2} \int_0^t \mathrm{d}\tau (t-\tau) \overline{\Delta E(\tau) \Delta
  E(0)}.\label{errprob}
\end{equation}
Since bosonic systems have purely Gaussian fluctuations, a similar
result is found within the spin-boson model~\cite{superdecoherence}.
The intuition of an error probability can be recovered in the limit of
small $\epsilon(N,t)$ since the fidelity decays as $F \approx 1 -
\epsilon(N,t)$. For correlated Gaussian phase noise the effective
error probability is then always proportional to
$N^2$~\cite{superdecoherence}. Therefore, initially negligible
correlated phase noise can lead to unexpectedly high error
probabilities as the size of the quantum register increases.

We can also obtain the Markovian and the static result for short
times by considering a correlation function of the form
$\overline{\Delta E(\tau) \Delta E(0)} = \overline{\Delta E^2}
\exp(-\gamma t)$ in the error probability (Eq.~\ref{errprob}).
Our noise model then leads to a fidelity
\begin{equation}
  F(t)=\frac{1}{2}(1+\exp(-\frac{1}{T_{2} \gamma} \{\exp(-\gamma t)+\gamma t
  -1\})), \nonumber
\end{equation}
where $T_2 = \hbar^2 \gamma / (N^2 \overline{\Delta E^2}) \propto
1/N^2$ corresponds to the decay time in the Markovian limit. For
$\gamma t \gg 1$, we recover the Markovian limit $F_{\mathrm{mark}}(t)
\cong 1/2 (1+\exp(-t/T_2))$ and for $\gamma t \ll 1$, the static
result for short times is $F_{\mathrm{stat}}(t) \cong 1/2
(1+\exp(-\frac{1}{2} (t/\tau)^2))$ with a charactistic time $\tau =
\hbar /(N \sqrt{\overline{\Delta E^2}}) \propto 1/N$.

Experimentally, we study correlated noise in an ion-trap quantum
processor. Our system consists of a string of $^{40}$Ca$^{+}$ ions
confined in a linear Paul trap where each ion represents a
qubit~\cite{expsetup}. The quantum information is encoded in the
S$_{1/2}$(m=$-1/2$) $\equiv \ket{1}$ ground state and the metastable
D$_{5/2}$(m=$-1/2$) $ \equiv \ket{0}$ state. Each experimental cycle
consists of three stages, (i) initializing the qubits and the center
of mass mode in a well defined state, (ii) performing the entangling
gate operation, and (iii) characterizing the quantum state. The qubits
are initialized by optical pumping into the S$_{1/2}$(m=$-1/2$) state
while the motion is brought to the ground state by Doppler cooling
followed by sideband cooling. Qubit manipulation is realized by a
series of laser pulses of equal intensity on all ions. The electronic
and vibrational states of the ion string are manipulated by setting
the frequency, duration, intensity, and phase of the pulses. Finally,
the state of the ion qubits is measured by scattering light at 397~nm
on the S$_{1/2} \leftrightarrow$~P$_{1/2}$ transition and detecting
the fluorescence with a photomultiplier tube (PMT). The camera
detection effectively corresponds to a measurement of each individual
qubit in the $\{|0\rangle,|1\rangle\}$ basis, while the PMT only
detects the number of ions being in $|0\rangle$ or $|1\rangle$.
Sufficient statistics is achieved by repeating each experiment 100
times for each setting.

In our system, GHZ states of the form ($\ket{0\ldots 0}+\ket{1\ldots
  1}$)/$\sqrt{2}$ are created from the state $\ket{1\ldots 1}$ through
a high-fidelity M{\o}lmer-S{\o}rensen (MS) entangling
interaction~\cite{MSgate,ghz_fault_tol}. Assessing the coherence,
fidelity, and entanglement of GHZ states is straightforward as the
density matrix ideally consists of only four elements: two diagonal
elements corresponding to the populations of $|0 \ldots 0 \rangle$ and
$|1 \ldots 1 \rangle$ as well as of two off-diagonal elements
corresponding to the relative coherence. The diagonal elements of the
density matrix $\rho$ are directly measured by fluorescence detection
and allow to infer the GHZ populations $P = \rho_{0 \ldots 0, 0 \ldots
  0} + \rho_{1 \ldots 1, 1 \ldots 1}$. The off-diagonal elements of
the density matrix are accessible via the observation of parity
oscillations~\cite{ghz_6_didi} as follows. After the GHZ-state is
generated, all qubits are collectively rotated by an operation
$\bigotimes_{j=1}^N \exp(i \frac{\pi}{4} \sigma^{(j)}_\phi)$ where
$\sigma^{(j)}_\phi = \sigma^{(j)}_x \cos\phi + \sigma^{(j)}_y \sin\phi
\;$ is defined by the corresponding Pauli operators on the $j$-th
qubit. By varying the phase $\phi$, we observe oscillations of the
parity
$\mathcal{P}=\mathcal{P}_\textrm{even}-\mathcal{P}_\textrm{odd}$ with
$\mathcal{P}_\textrm{even/odd}$ corresponding to the probability of
finding the state with an even/odd number of excitations. The
amplitude of these oscillations directly gives the coherence $C =
|\rho_{0\ldots 0, 1 \ldots 1}|+|\rho_{1\ldots 1, 0 \ldots 0}|$ of the
state. The fidelity of the GHZ state is then given by $F=(P+C)/2$,
where a $F>50\%$ implies genuine N-particle
entanglement~\cite{ghz_6_didi}. States with a $F<50\%$ can still be
genuinely N-particle entangled if they satisfy other criteria. We
apply the criteria defined in
Ref.~\cite{ghz_threshold_wolfgang,ghz_crit_otfried} which can be used
in conjunction with the procedure described above. One criterion tests
multipartite distillability (i.e. N-particle entanglement can be
distilled from many copies of this state)
\cite{ghz_threshold_wolfgang}, while a more stringent criterion proves
genuine N-particle entanglement \cite{ghz_crit_otfried}.

\begin{figure}[ht!]
 \begin{center}
  \includegraphics[width=\columnwidth]{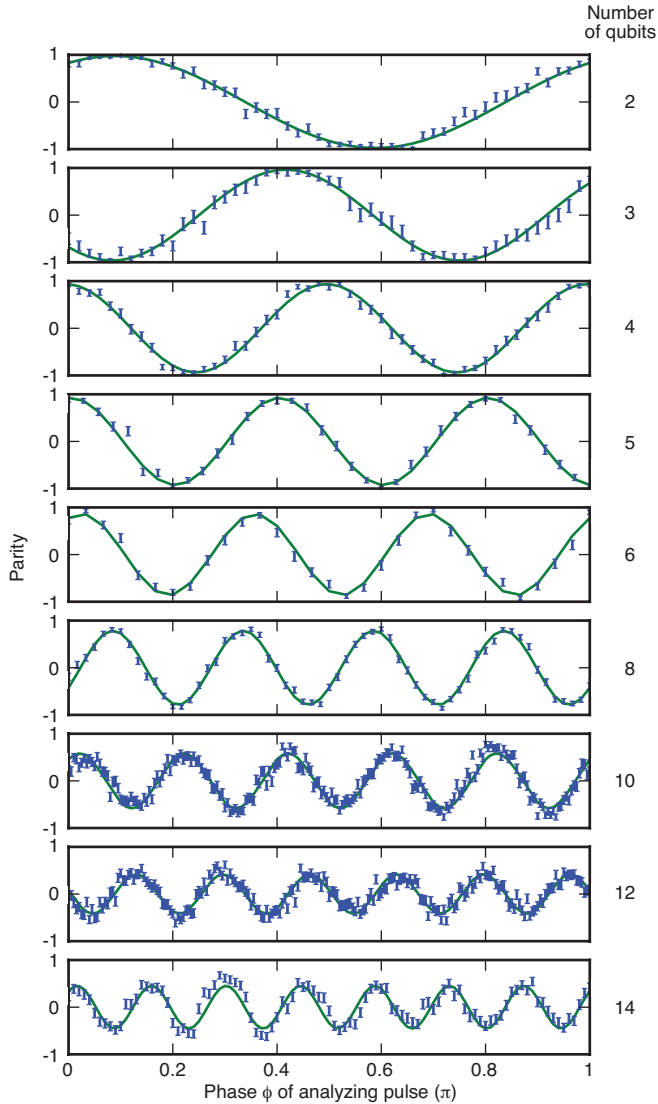}
  \caption{ Parity oscillations observed on \{2,3,4,5,6,8,10,12,14\}-qubit
    GHZ states.}
  \label{fig:parity_oscillations}
 \end{center}
\end{figure}

We have experimentally prepared GHZ states of \{2-6,8,10,12,14\} ions and
achieved the populations, coherences, and fidelities shown in Table I. The
observed parity oscillations are shown in Fig.~\ref{fig:parity_oscillations}.
Although N-particle distillability can be inferred from the criterion in
Ref.~\cite{ghz_threshold_wolfgang} by many standard deviations, according to
the criteria in Ref.~\cite{ghz_crit_otfried} the obtained data support genuine
N-particle entanglement for 14 qubits with a confidence of 76\%. The 12-qubit
state is likely not fully entangled. The Poissonian statistics of the PMT
fluorescence data is accounted for by a data analysis based on Bayesian
inference~\cite{bayes_photoncounting}.


\begin{figure*}
  \includegraphics{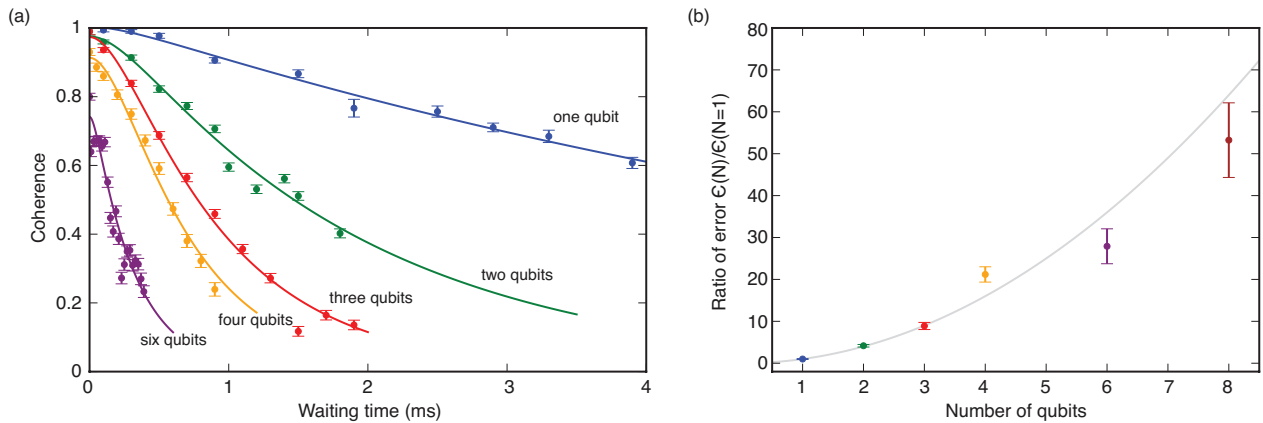}
  \caption{Coherence decay and relative error probability $\epsilon(N)$ of GHZ
    states. (a) Remaining coherence as a function of time for a single qubit
    (blue) and GHZ states of 2 (green), 3 (red), 4 (orange), and 6 (purple)
    qubits. (b) The observed relative error probability is consistent with a
    scaling behavior proportional to $N^2$ as indicated by the gray line. The
    coherence of an $N$-qubit GHZ state then decays by a factor $N^2$ faster
    than the coherence of a single qubit.}
  \label{fig:relative_noise}
\end{figure*}

The coherence of GHZ states as a function of time is investigated by adding
waiting times between creation and coherence investigation. The observed
coherence decay, equivalent to an error probability, is directly compared with
that of a single qubit, ideally yielding a relative error probability
$\epsilon(N)=\epsilon(N,t)/\epsilon(1,t)=N^2$. The obtained data
(Fig.~\ref{fig:relative_noise}) is consistent with an $N^{2.0(1)}$ scaling law,
in full agreement with predictions for correlated Gaussian noise. In other
words, the coherence of an $N$-qubit GHZ state decays by a factor $N^2$ faster
than for a single qubit. The scaling is here explored with up to 8 qubits
because for more qubits the quality of the entangling gate is currently too
sensitive to slow drifts in the experimental apparatus.

As several systems experience correlated noise, this superdecoherence
will eventually limit the overall performance of large-scale quantum
registers (unless qubits are encoded in noise-insensitive
subspaces~\cite{dfsbasics1,dfsbasics2}). In our experiment, the noise
affecting the quantum register is mainly caused by fluctuations of the
homogeneous magnetic field due to a varying current in the field
generating coils. By decreasing this noise, the single-qubit coherence
time improved ten-fold from $8(1)$~ms to $95(7)$~ms. Such
coherence time is approximately a factor of 1000 longer than the gate
time of the MS interaction of approximately 100~$\mu s$. This long
coherence time would, in principle, enable the implementation of
algorithms with 10 and more qubits. In the presence of correlated
noise, however, this $N^2$ scaling can potentially be the main
limitation for several experiments. A correlated phase-noise
environment with a single-qubit characteristic error probability of
only 0.01 leads to a 10-qubit GHZ-state relative error probability
$\epsilon(N=10) = 0.01 \times 10^2 \approx 1$; most of the state's
phase information is then lost.

We verify that correlated phase noise is dominant in our experiment by
preparing a state which is insensitive to this noise.  We create the state
$\ket{00001111}+\ket{11110000}$)/$\sqrt{2}$, which is locally equivalent to an
8-qubit GHZ state. This state is realized by an MS interaction starting from the
state $\ket{00001111}$. Its coherence properties are investigated as above
using a local transformation into a GHZ state. The state shows a coherence time
of 324(42)~ms. This result is consistent with a lifetime-limited quantum state
with an effective lifetime of one fourth that of a single qubit of 1.17~s. In our
apparatus, this extension of the coherence time relative to the GHZ state (or
even the single-qubit case) can only be explained by correlated noise affecting
the entire quantum register. Employing such insensitive states will therefore
be crucial for large-scale quantum information processing affected by
correlated phase noise.

Being able to efficiently generate entangled quantum states involving
10 and more qubits opens a new range of applications. Our system
represents the basic building block for quantum simulation
experiments~\cite{qsim} to investigate complex mechanism such as the
magnetic sense of birds~\cite{magnetic_bird_sensor}, to perform
exponentially compressed spin-chain
simulations~\cite{compressable_isingxy}, and to better understand
cosmology and space-time~\cite{universe_via_10ions}. It may serve as a
very well-controlled testbed for fundamental questions in quantum
physics such as the investigation of the cross-over from
superpositions in quantum systems to defined states in macroscopic
systems with GHZ states~\cite{macro_realism}.


In conclusion, we have analyzed the decay of GHZ-states in an ion-trap
based quantum computer. We find a dependency that scales quadratically
with the number of qubits and thus shows superdecoherence . This
mechanism is present in every other experiment that relies on a phase
reference for performing quantum information processing with
energetically non-degenerate qubits. Superdecoherence may especially
affect quantum metrology based on GHZ states. We achieve coherence
times of about 100~ms on an optical qubit which is a factor of 1000
longer than an entangling gate operation in the same system. Using a
single-step entangling gate based on the ideas of M{\o}lmer and
S{\o}rensen, we generate genuine multiparticle entangled states with
up to 14 qubits. The employed techniques represent an encouraging
building block for upcoming realizations of advanced quantum
computation and quantum simulations with more than 10 qubits.

We thank O.~G\"uhne for helpful discussions. We gratefully
acknowledge support by the Austrian Science Fund (FWF), by the
European Commission (AQUTE, STREP project MICROTRAP), IARPA, the CIFAR
JFA, NSERC, and by the Institut f\"ur Quanteninformation GmbH. This
material is based upon work supported in part by the U. S. Army
Research Office. J.T.B. acknowledges support by a Marie Curie
International Incoming Fellowship within the 7th European Community
Framework Programme.


\end{document}